\documentclass[showpacs,aps]{revtex4}
\usepackage{epsf}    
\begin{document}
\newcommand{\rp}{\right)}
\newcommand{\lp}{\left(}     

\title{Causalities of the Taiwan Stock Market}

\author{Julian Juhi-Lian Ting}
\email{jlting@chez.com}
\homepage{http://www.chez.com/jlting}
\affiliation{No.38, Lane 93, Sec.2,
Leou-Chuan E. Rd., Taichung, Taiwan 40312, ROC.}
\date{\today}

\begin{abstract}
Volatility, fitting with
first order Landau expansion, stationarity, and causality
of the Taiwan stock market (TAIEX)
are investigated based on daily records.   
Instead of consensuses that consider stock market
index change as a random time series we propose
the market change
as a dual time series consists of the index and the corresponding volume.
Therefore, causalities between these two time series are investigated.
\end{abstract}

\maketitle

\section{Introduction}
Physicists are interested in studying stock markets as complex systems.
Almost all questions asked can be summarized as searching for price formation theories.
Previous researches shown the distribution of price
change has pronounced tail distribution in contrast to
Gaussian distribution expected. Furthermore, the auto-correlation of
price change decays exponentially with a characteristic time scale 
around 5 minutes. 
Stock crash or rally have also been identified by physicists as a
kind of herd behaviour\cite{EZ}.
                           
Johansen and Sornette considered fitting most
stock markets for the bubbles using Landau expansions of the index\cite{JS1}.
They showed evidence that market crashes as well as large corrections are
preceded by speculative bubbles with two main characteristics:
a power law acceleration of the market price decorated with log-periodic
oscillations. 
For most markets
the log-frequency $\omega/2\pi$ is close to unity. 
However, most data analysis were done for the index of more mature market 
like S \& P 500.
Not all market indexes are defined in the same way.
Could different weighting methods reach the same conclusion?
Will government intervention play a role in the conclusion?
Johansen and Sornette further  found emergent markets
have larger fluctuations.
They extended the expansion up to third order and successfully
predicted Nikkei raise in the year 2000\cite{JS2}.
However, why do these fluctuations exist?
The log-periodic oscillation appeared in a wide class of out of equilibrium dynamic systems, like ruptures in heterogeneous media,
historic analysis of earthquakes data, and world population.
Canessa tried to establish universality for the exponents 
from a renormalization 
group theory\cite{EC}, and used a stochastic theory
to show that the log-periodicities are a consequence of transient clusters introduced by an entropy-like term.
As entropy in thermodynamics corresponds to the information in an
information theory.
The possibility to arrive at the log-periodic oscillation therefore suggests the
log-periodic oscillation is a consequence of 
information exchange between different species of a large system\cite{EC2}.

The effect of volume is less analyzed in the literature.
Volume is  a measure of market liquidity while index means the price.
Gopikrishnan {\it et al.}\cite{GPGS} analyzed the statistical properties
of number of shares traded of a particular stock at  a given time interval
from an empirical rule saying that it takes volume to push the index.
Bonanno {\it et al.}\cite{BLM} also analyzed the number of shares
traded of selected stocks and find a power spectrum of approximately
$1/f$.

In the present work we considered the volume effects of Taiwan stock market
(TAIEX). In particular, the cause-effect relation between the volume time series
and the index time series is analyzed.
Taiwan stock market is one of the largest emerging markets. 
Johansen and Sornette did not studied it
because of availability of trading information.
In the present work we tried to make up this missing piece.

When people talk about the volume involved,
sometimes they mean the number of shares
traded, sometimes the amount of money involved.
We consider money flow to be more important than share number flow
if the whole market is considered, since the person
or institute involved should have fixed amount of money. 

\section{Basic Properties}
Taiwan has two stock markets. TAIEX is the major market\cite{TAIEX}.
TAIEX definition is taken the yearly average of 1966 as 100.
All stocks traded are taken into account.

We analyzed our daily data of TAIEX from January 3,
1991 to December 30, 2000, which include 2814 trading days for a ten years period.
Intra-day time is treated as continuous.
Within our data time, several significant events can be identified. 
In August 26, 1997, the market reached a local peak 10116.84. 
The highest index happened at February 17, 2000 with an index 10202.2. 
The lowest point is at January 7, 1993 with an index 3135.56.

\begin{figure}[bth]
\vskip 0.1cm
\epsfxsize=8.5cm\epsfbox{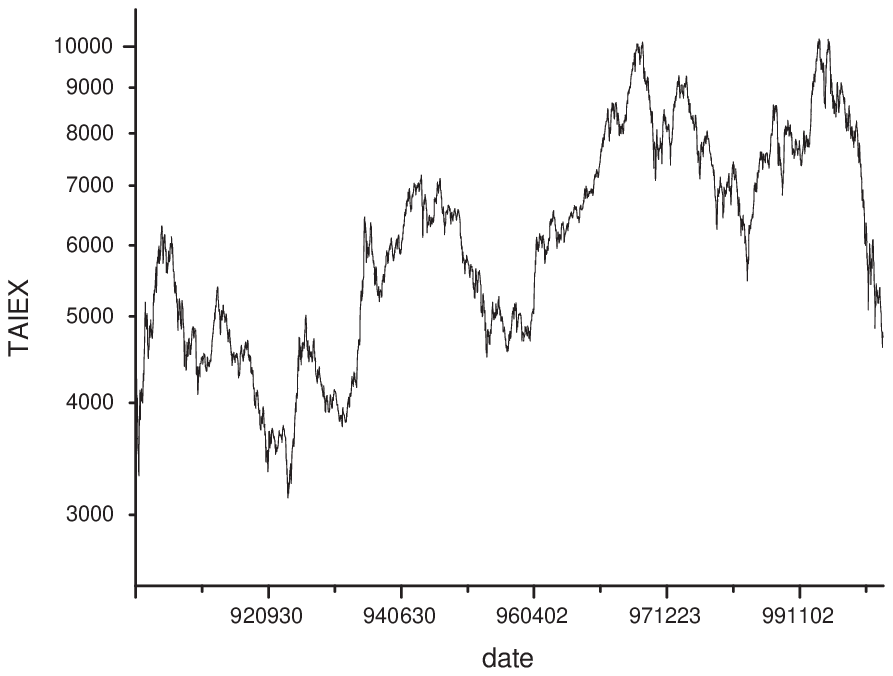}
\epsfxsize=8.5cm\epsfbox{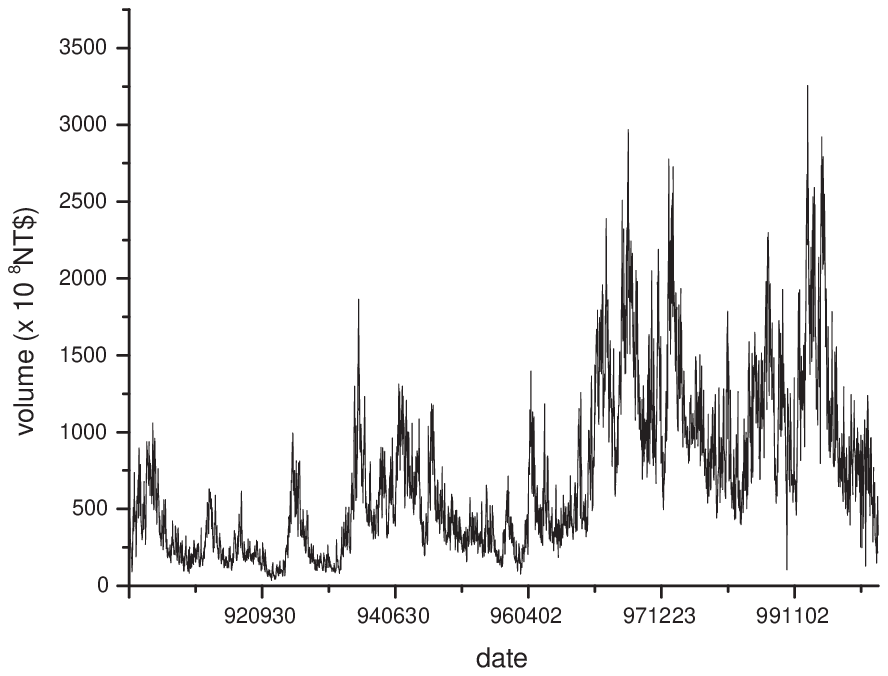}
\vskip 0.1cm
\caption{Histories of TAIEX daily variations and the corresponding volumes.}
\label{taiex}
\end{figure}

In order to compare Taiwan stock market with other countries we
need to know some basic properties studied in other markets.
The first properties are of course daily index and volume histories.
They are plotted in Fig.(\ref{taiex}).

\begin{figure}[bth]
\vskip 0.1cm
\epsfxsize=8.5cm\epsfbox{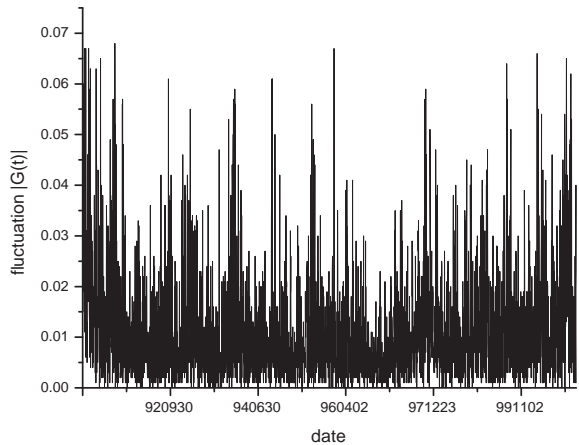}
\vskip 0.1cm
\caption{Daily fluctuations amplitude, $|G(t)|$.}
\label{fluctuation}
\end{figure}

One of the most studied properties is volatility.
It is found to be a good measure for long time behaviour
and risk\cite{LGCMPS}. It is also the key input parameter for all
option pricing models.
Many different quantitative definitions of
volatility are used in the literature. 
Following Liu {\it et al.}
we define the price change $G(t)$ as the change in
the logarithm of the index,
\begin{equation}
G(t)\equiv\ln Z(t+\Delta t) -\ln Z(t) \cong \frac{Z(t+\Delta t)-Z(t)}{Z(t)}\;,
\label{eq:gt}
\end{equation}
where $\Delta t$ is the sampling time interval. 
The absolute value of $G(t)$ describes the amplitude of the fluctuation,
as shown in Fig.~(\ref{fluctuation}).
The large values of
$|G(t)|$ correspond to the crashes and big rallies.
None  of the $|G(t)|$ ever be larger than $0.07$ because government
regulation limited maximum  daily changes to $7\%$.

\begin{figure}[bht]
\vskip 0.1cm
\epsfxsize=8.5cm\epsfbox{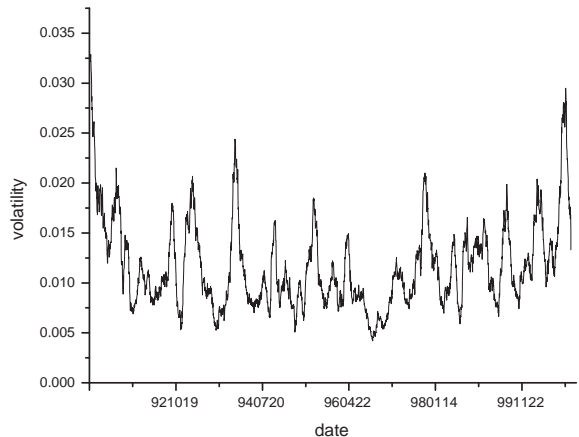}
\vskip 0.1cm
\caption{Volatility history over a time window $T = 30 $ days. Dates are labeled
at the centers of the average windows.}
\label{tvhis}
\end{figure}

We define the volatility as the average of $|G(t)|$ over a time window
$T=n\cdot\Delta t$, i.e.,
\begin{equation}
V_T(t)\equiv{1\over n}\sum_{t'=t}^{t+n-1}|G(t')|\;,
\label{defV}
\end{equation}
in which $n$ is the moving window size. 
Fig.~(\ref{tvhis})
shows the calculated volatility $V_T(t)$ for a large
averaging window $T=30\,$days.

\begin{figure}[hbt]
\vskip 0.1cm
\epsfxsize=8.5cm\epsfbox{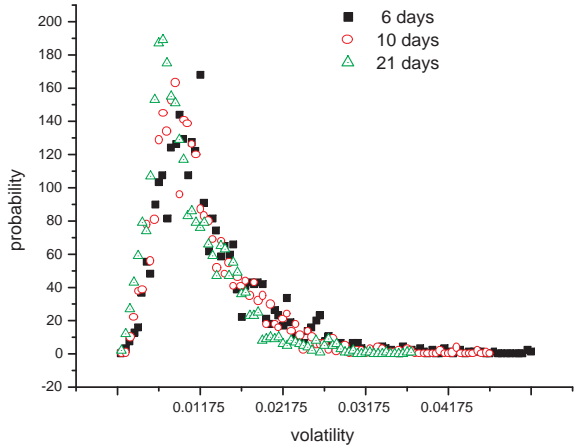}
\vskip 0.1cm
\caption{Volatility distribution for various time windows with
$\Delta t = 30$ min.}
\label{prob}
\end{figure}

\begin{figure}[hbt]
\vskip 0.1cm
\epsfxsize=8.5cm\epsfbox{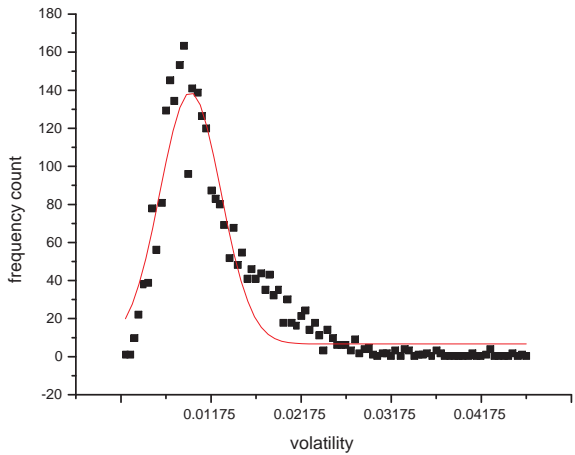}
\vskip 0.1cm
\caption{Gaussian fit for the volatility distribution for $T=10 $ days and
$\Delta t = 1$ days. }
\label{fit10}
\end{figure}

Fig.~(\ref{prob}) shows the probability density function $P(V_T)$
of the volatility for several values of $T$ with $\Delta t=1\,$day. 
The 10-day  data is further fitted with Gaussian distribution in Fig.~(\ref{fit10}).
Apparently the data distribution deviates from the Gaussian curve
in many places. The right hand side heel has much higher probability
than the Gaussian fit.

\begin{figure}[hbt]
\vskip 0.1cm
\epsfxsize=8.5cm\epsfbox{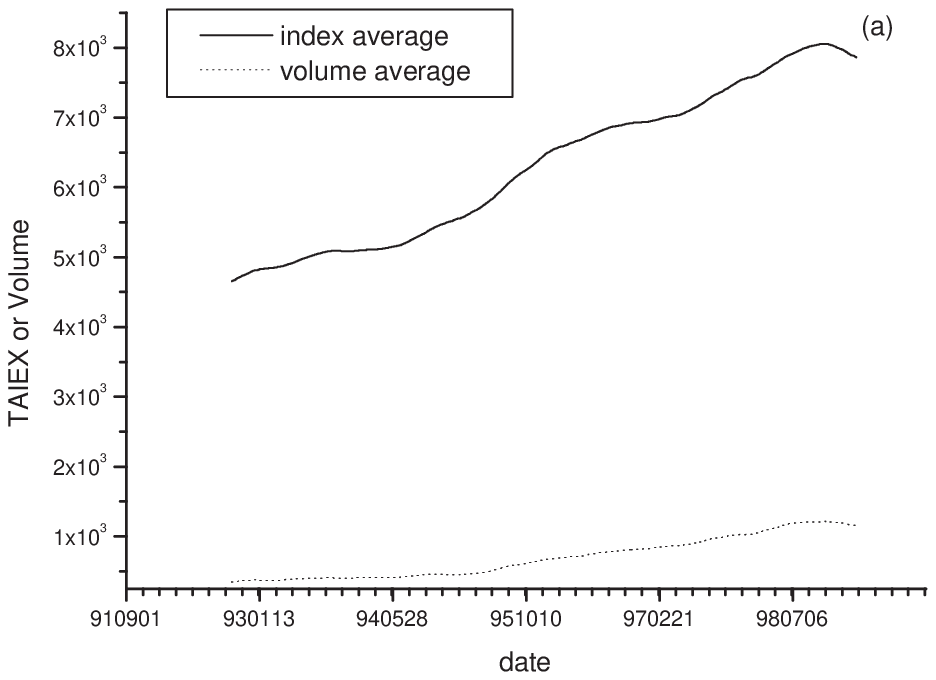}
\epsfxsize=8.5cm\epsfbox{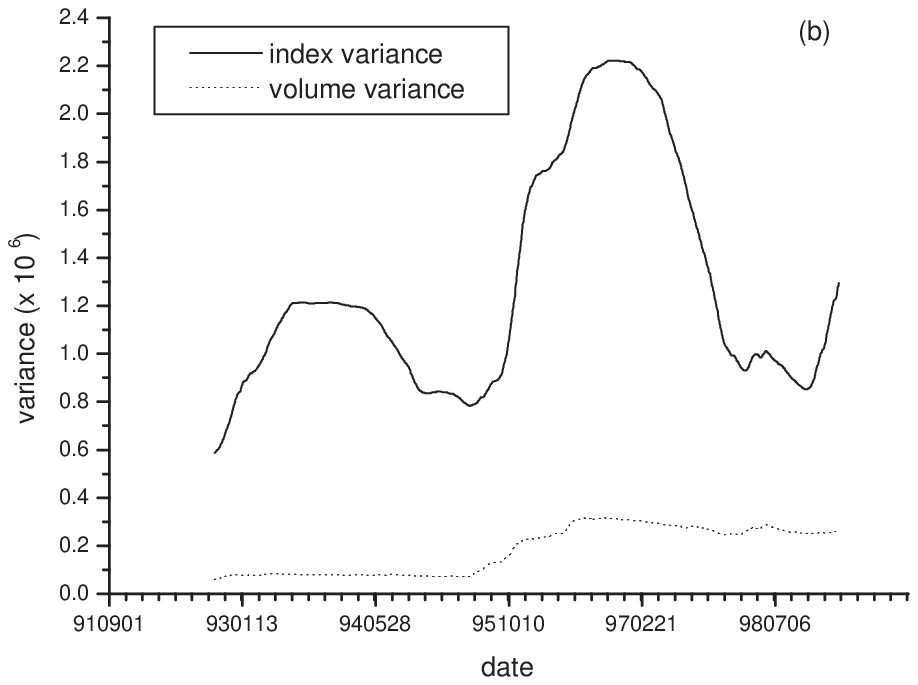}
\vskip 0.1cm
\caption{Stationarity test for TAIEX with a window size $T=1000 $ days for (a) average of the index and the corresponding volume (b) variance of the index and the corresponding volume. The window is labeled by its central date.}
\label{station}
\end{figure}

An important property need to be investigated for any time series analysis is its
stationarity. 
Most empirical analysis assumes the time series to be
stationary. A time series is stationary if its mean value and its variance do not vary systematically over time\cite{DNG}. 
For TAIEX we choose a window of 1000 trading days, say, and calculate the corresponding mean value and variance. The result is plotted in Fig.~(\ref{station}). We find the averages of both index and volume 
have clear upward trend, while the variances of index oscillate. 
The volume average and its variance curves have a turning point
around the year 1995, which marks a fundamental change of
Taiwan economics.
The index curve is further fitted with a linear curve. The slope is
$1.6$, which means if an investor holds a index portfolio for over 1000 days her expectation to profit is 1.6 point/day. 

\begin{figure}[hbt]
\vskip 0.1cm
\epsfxsize=8.5cm\epsfbox{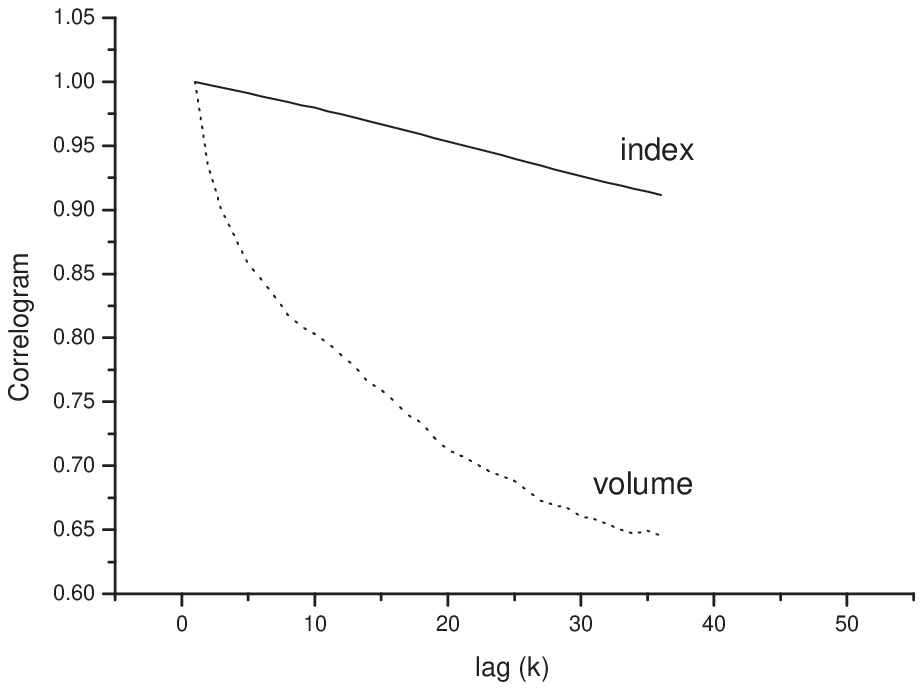}
\vskip 0.1cm
\caption{Correlogram plot for both index and the corresponding volume time series.}
\label{correl}
\end{figure}

Another test for stationarity is based on sample 
autocorrelation function,
$\rho_k=\gamma_k / \gamma_0$, in which
$\gamma_k={{\sum (Y_t-\bar{Y}) (Y_{t+k}-\bar{Y}) }}$
is the covariance at lag $k$. If $\rho_k$ decays very fast with increasing $k$,
the time series is stationary. The resulting correlogram is plotted in Fig.(\ref{correl}). Apparently, 
the index time series is more stochastic than the volume time series. The results of Fig.(\ref{station}) and Fig.(\ref{correl}) are consistent.

\section{crashes}
\label{crashes}
In this section,
we follow Johansen and Sornette
to fit the index changes and obtain future trends prediction.
However, our analysis differs from theirs in labeling the date.
They convert all dates into decimal numbers. We labeled only by trading dates.
Our labeling method will be more reliable in dating, since
trading days spread irregularly amount years.

Johansen and Sornette wrote down the first, second, and third
order Landau expansion of the index starting or ending at
the historical high $t_c$. 
The first order
expansion reads:
\begin{equation}
p_1\lp t\rp \approx A + B \tau ^{\alpha}
+C \tau^\alpha \cos\left[ \omega\ln \lp \tau \rp +\phi \right].
\end{equation}
in which $\tau = | t - t_c |$ is the absolute value of time difference
from the critical time of the highest index $t_c$.
The parameter $\phi$
correspond to the time unit used,
while
the parameter $A , B ,$and $C $ are units determined by the index
as well as the historical period.
Only $t_c, \alpha ,$ and $ \omega$ are the key parameters:
$\alpha $ stands for the power law acceleration, while 
$\omega$ quantifies for the log-periodic oscillation.
Numerically, $A$ is roughly the index at $t_c$.

Our fitting start from 
January 7, 1993 to February 17, 2000.
\begin{figure}[hbt]
\vskip 0.1cm
\epsfxsize=8.5cm\epsfbox{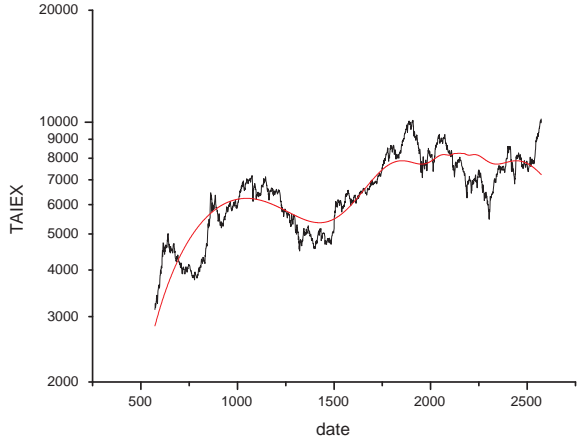}
\vskip 0.1cm
\caption{TAIEX fit by first order Landau expansion of the index starting from
January 7,1993 to February 17,2000.
Fitted parameters are:
$t_c=2147.999$, $\alpha=1.276$, $\omega= 11.013$, $\phi=7.037$,
$A=8250.314$, $B=-0.49$, and $C=-0.236$.
}
\label{bubble}
\end{figure}
The result is shown in Fig.(\ref{bubble}).
The fitting does not appears good, since it is only a first order approximation
while the fitting period is about 7 years.
There exits a fitting solution with $t_c$ far beyond the highest index
point. However, we cannot take that solution because it is obviously incorrect.
The fitted $t_c$ has to come before the highest point.

\begin{figure}[hbt]
\vskip 0.1cm
\epsfxsize=8.5cm\epsfbox{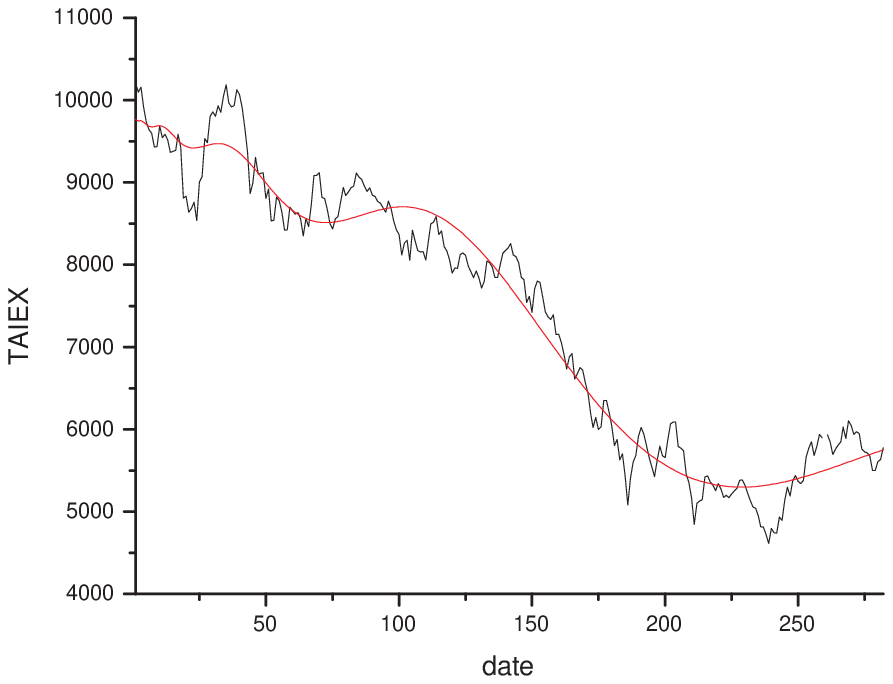}
\vskip 0.1cm
\caption{TAIEX fit by first order Landau expansion of the index starting from
February 17, 2000.  Fitted parameters are:
$t_c=-0.491$, $\alpha=1.106$, $\omega= 12.650$, $\phi=5.558$,
$A=9776.675$, $B=-9.079$, and $C=2.926$.
}
\label{future}
\end{figure}

A theory will only be useful if it can also predict for the future.
We used the same equation to fit from February 17,2000 in Fig.(\ref{future}).
We find the future is promising.
However, in both cases the $\omega$ were found to be much higher than the
previous authors'.

\section{Causality Test}

It is well known to the practitioner, as an empirical rule, that
volume changes come before price change.
A rule of thumb is that higher index will come after huge volume,
and if the volume is reaching a minimum the bottom of the index
will not be far.
However, how much ahead? What kinds of correlation do they posses?
Why is there a lag?
This kind of dual series with one influencing the
other has been analyzed in the econometric literature for GNP (gross national product)-money supply relation, for example\cite{DNG,GSM}.
It is called causality as in the literature of physics.
We have already known our time series are non-stationary.
In order for the causality test to be meaningful, the two time series have
to be cointegrated, i.e. their wave-lengths of variation have to be of the same order.
Fig.(\ref{station})(b) and Fig.(\ref{correl}) show their time scale
differs but not far.
                    
In the econometrics literature a linear functional form is
used for the regression of these two time series.
Furthermore, because the future cannot cause the past,
we can write down a functional form for the future
index, $I_t$ and the future trading volume, $V_t$, as
\begin{eqnarray}
I_t&=\sum_i a_i V_{t-i} + \sum_j b_j I_{t-j} + u_t \;\\
V_t&=\sum_i l_i V_{t-i} + \sum_j k_j I_{t-j} + v_t \;.
\label{causality}
\end{eqnarray}

\begin{figure}[hbt]
\vskip 0.1cm
\epsfxsize=8.5cm\epsfbox{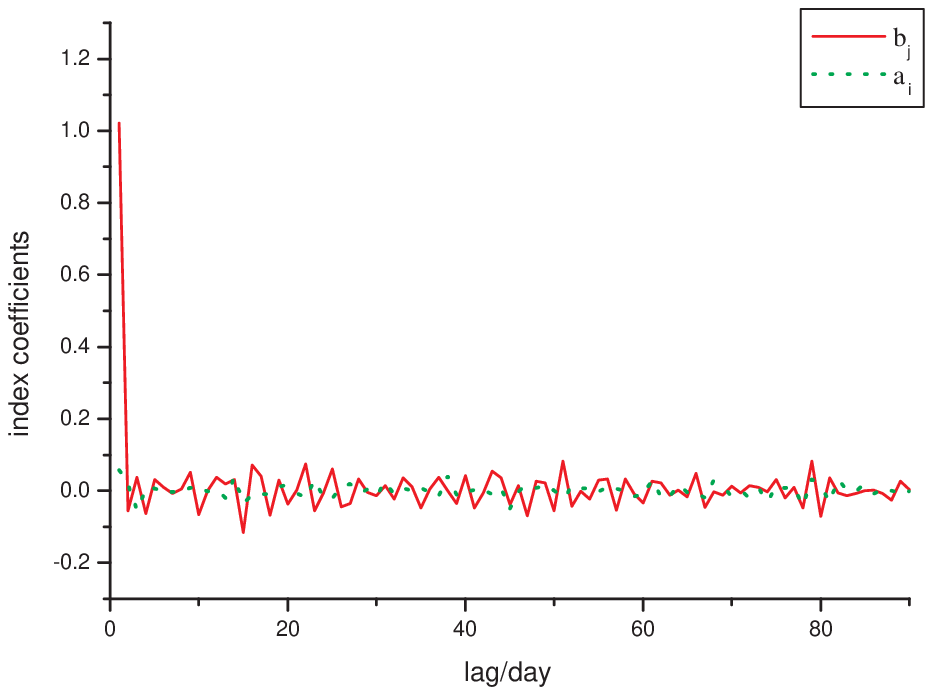}
\epsfxsize=8.5cm\epsfbox{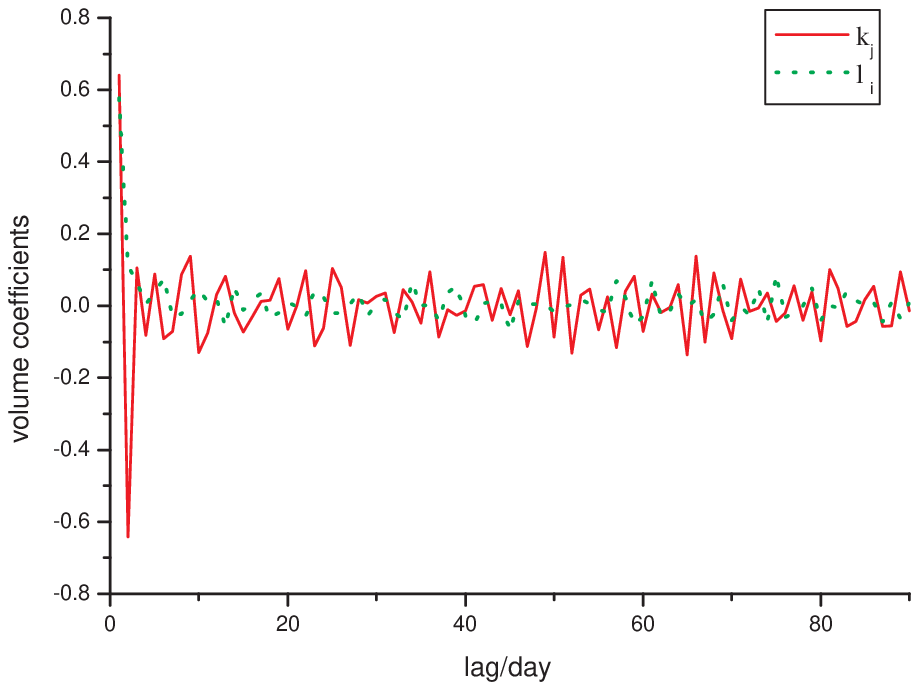}
\vskip 0.1cm
\caption{Causality coefficients for index and for volume.}
\label{cause}
\end{figure}
\begin{figure}[hbt]
\vskip 0.1cm
\epsfxsize=8.5cm\epsfbox{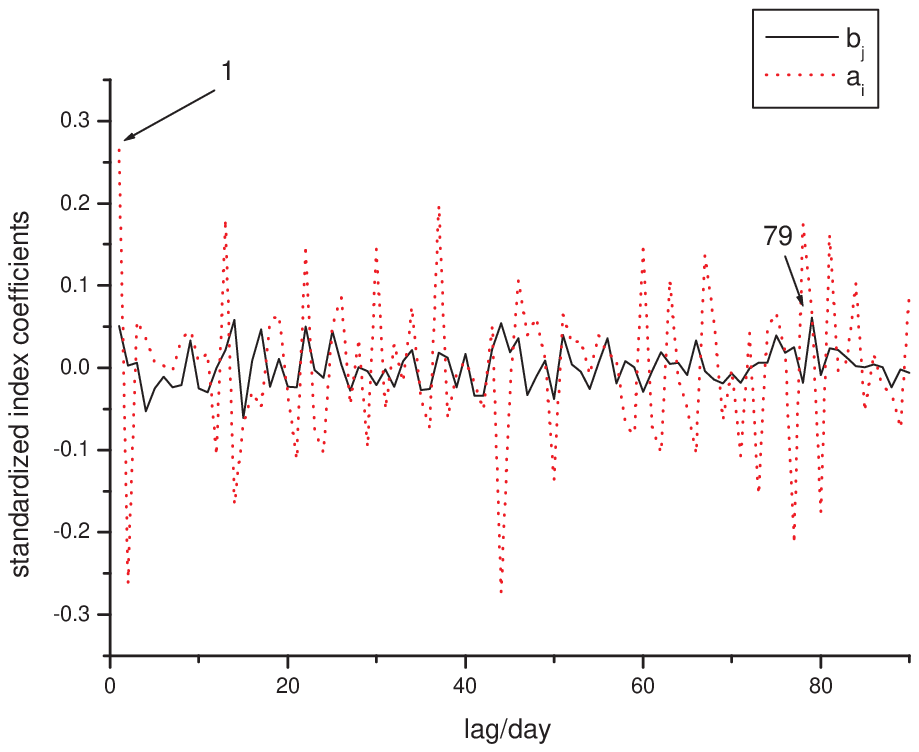}
\epsfxsize=8.5cm\epsfbox{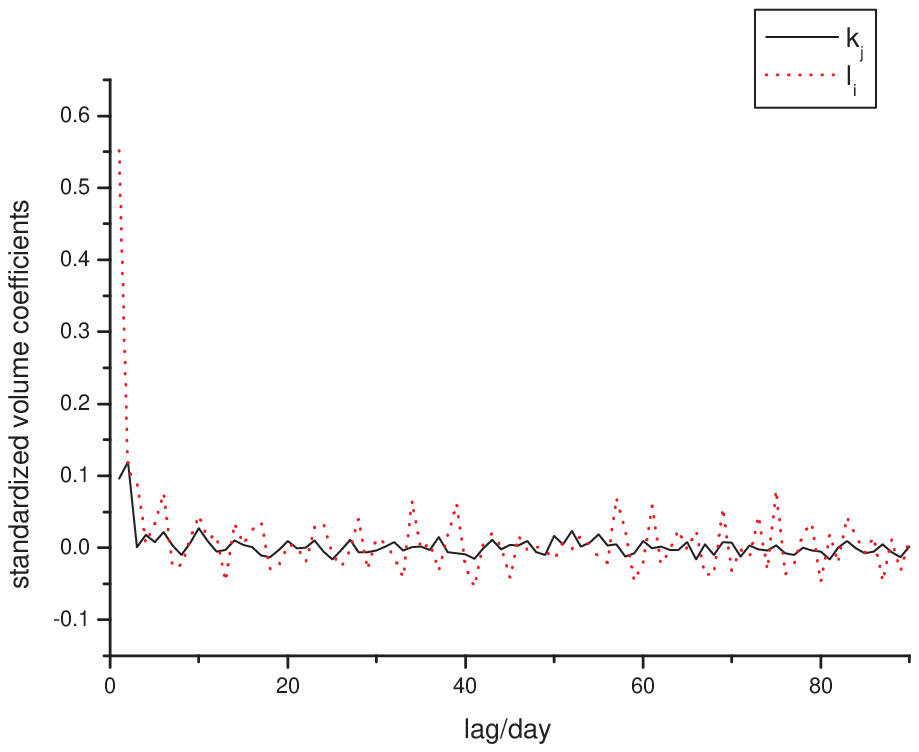}
\vskip 0.1cm
\caption{Causality coefficients for index difference and for the corresponding volume.}
\label{causdiff}
\end{figure}

We fitted for 90-days from the history using the above   functional form.
The result is plotted in Fig.(\ref{cause}). 
Both the index fitting and the volume fitting show stronger dependence on the index history than on the volume history. 
However, only the previous trading day shows significance.
Another plot is made for the index difference and the corresponding volume for the future
index difference, $I_t$ and the future trading volume, $V_t$, as
\begin{eqnarray}
\delta I_t&=\sum_i a_i V_{t-i} + \sum_j b_j \delta I_{t-j} + u_t \;\\
V_t&=\sum_i l_i V_{t-i} + \sum_j k_j \delta I_{t-j} + v_t \;.
\label{caudiff}
\end{eqnarray}
The result is plotted in Fig.(\ref{causdiff}). 
The figure suggests the index difference
time series indeed received slightly higher
influence from
the previous index difference and the previous volume. 
However, these effects are pretty small.
Furthermore, these peaks will be changed with the truncation of the time series.

\section{conclusion}
We tried to find the correlation between  the volume
of transactions involved and stock market fluctuations beyond simply
considering temporal price series, using Taiwan stock market
as an example in the present work. 

We find TAIEX volatility spread relatively uniformly in the past ten years, 
which can also be found in the stationarity test.
The distribution of price change shows fat-heel behaviour.

A slightly higher influence coefficient is found in the index difference
fitting with previous index and previous volume.
Johansen and Sornette's result suggests us to try regression with log-periodic functional forms in the future. 
However, it does not seem meaningful to go beyond 90-days for the time lag, since companies will have their seasonal reports by that time.


\end{document}